\documentclass[showpacs,twocolumn,amsmath,pre,superscriptaddress,nofootinbib]{revtex4}
\usepackage{graphicx}
\usepackage{amsmath}

\usepackage{epsfig}
\usepackage{graphics}
\usepackage{latexsym}
\usepackage{amsfonts}
\usepackage{amssymb}
\usepackage{color}

\usepackage{plain}

\begin{document}

\title{The $T=0$ RFIM on a Bethe lattice: correlation functions along the hysteresis loop}

\author{Xavier Illa}
\affiliation{Department of Applied Physics, Aalto University, PO Box 14100, Aalto 00076, Finland}
\email{xavier.illa@aalto.fi}

\author{Martin Luc Rosinberg}
\affiliation{Laboratoire  de  Physique  Th\'eorique  de  la  Mati\`ere
Condens\'ee,  CNRS-UMR 7600,  Universit\'e  Pierre et  Marie Curie,  4
place Jussieu, 75252 Paris Cedex 05, France}
\email{mlr@lptmc.jussieu.fr}

\begin{abstract}
We consider the Gaussian random field Ising model (RFIM) on the Bethe lattice at zero temperature in the presence of a uniform external field and derive the exact  expressions of the two-point spin-spin and spin-random field correlation functions along the saturation hysteresis loop. To complete the analytical description and  suggest possible approximations for the RFIM on Euclidian lattices we also compute the corresponding direct correlation functions (or proper vertices) and show that they decay rapidly with the distance in the weak-coupling/large disorder regime; their range, however, is not  limited to the nearest-neighbor distance.
\end{abstract}


\pacs{75.60.Ej, 75.10.Nr, 05.50.+q} 
\maketitle

\def\be{\begin{equation}}
\def\ee{\end{equation}}
\def\bea{\begin{align}}
\def\eea{\end{align}}

\section{Introduction}

The random field Ising model (RFIM) at zero temperature is a simple prototype of a class of  disordered systems 
(such as random magnets, martensitic materials, fluids in porous solids,...) that exhibit hysteretic and jerky 
behavior when slowly driven by an external field\cite{SDP2006}.  The most interesting feature of the model which has been the subject of extensive analytical and numerical studies is the existence of a disorder-induced nonequilibrium phase transition between two different regimes of avalanches\cite{SDKKRS1993}. This transition manifests through a change in the shape of the magnetization hysteresis loop that evolves from continuous to discontinuous as the disorder strength is reduced.  The discontinuity is associated to a macroscopic avalanche involving a finite fraction of the spins in the thermodynamic limit. As shown recently, this type of mechanism plausibly explains the hysteretic behavior of $^4$He adsorbed in high porosity silica aerogels\cite{DKRT2005,B2008}. Interestingly,  this nontrivial behavior is already present on the Bethe lattice  ({\it i.e.} the infinite Cayley tree) where a fully analytical characterization of the major and minor hysteresis loops, the avalanche size distribution, and other quantities can be obtained thanks to the tree topology\cite{DSS1997,CGZ2002,ABCDDMZ2005,IOV2005,OS2010}. In this case, the out-of-equilibrium phase transition occurs when the coordination number $z\ge4$ and is described by a traditional saddle-node transition in the self-consistent field equation\cite{DSS1997} (which makes the critical behavior the same as that for the infinite-range mean-field model).
In this work we extend the analytical description to the spin-spin and spin-random field correlation (or Green's) functions along the hysteresis loop, using the fact that correlations on a tree-like graph have a one-dimensional character. For a Gaussian distribution of the random fields, the spin-random field correlation function is also related to the slope of the magnetization curve through a `susceptibility' sum-rule. The motivation for this calculation is twofold. First, on the theoretical side,  Green's functions (or, better, their matrix inverse, the so-called direct correlation functions in liquid state theory or proper vertices in field-theoretic language) may be  used as the building blocks of approximate theories, as illustrated by the recent computation of the hysteresis loop in the three-dimensional soft-spin random field model\cite{RT2010}.  Exact results, even for simple models, may give some insight of the actual structure of these functions. Secondly, on the experimental side, scattering methods are now frequently combined with other standard probes (response to an applied field or thermodynamic measurements) for extracting information on the structure and the dynamics of  systems with quenched randomness (see {\it e.g.} Ref. \cite{H2004} in the case of fluids adsorbed in porous solids). Knowing the structure of the correlation functions can thus make easier the interpretation of the scattered intensity\cite{DKRT2006}. 

The outline of the paper is as follows. In section II, we define the model and give the expressions of the correlation functions, first for the one-dimensional chain (correcting the result obtained in Ref.\cite{KFS2000}), and then generalizing to the Bethe lattice (the detailed calculations are presented in Appendices A and B).  Analytical  predictions are compared to simulations performed on regular random graphs. In section III, we compute the corresponding direct correlation functions. We then conclude.

\section{Model and correlation functions}

The RFIM is defined by the following Hamiltonian
\begin{align}
{\cal H}=-J\sum_{<ij>} S_iS_j +\sum_i (H+h_i)S_i
\end{align}
where the $N$ spins $S_i=\pm 1$ are placed on the vertices of a Bethe lattice with coordination number $z$. The first sum is restricted to nearest-neighbors (n.n.) pairs and $J>0$. $H$ is a uniform external field and  the fields $\{h_i\}$ are  random variables drawn independently from a Gaussian distribution $\rho(h)=\exp(-h^2/2\Delta)/\sqrt{2\pi \Delta}$ with the variance $\Delta$ measuring the  strength of disorder. 

The relaxation dynamics is the $T=0$ limit of the Glauber dynamics and consists in aligning the spins with their local effective field at each time step\cite{SDKKRS1993},
\begin{align}
\label{Eqrelax}
S_i =\mbox{sgn}(f_i)	
\end{align}
where
\begin{align}
\label{Eqlocfield}
f_i =-\frac{\partial {\cal H}}{\partial S_i}=J\sum_{j/i}S_j+H+h_i
\end{align}
and the sum runs over the $z$ nearest neighbors of site $i$. The dynamics thus proceeds via a series of  avalanches which stop when a metastable state is reached, {\it i.e.}, when all spins satisfy Eq. (\ref{Eqrelax}). The saturation hysteresis loop is  obtained by adiabatically ramping $H$ from $-\infty$ to $+\infty$ and back. Thanks to the tree topology of the Bethe lattice and the abelian property of the dynamics ({\it e.g.} the fact that the metastable state after an avalanche does not depend on the order in which the spins flip), the shape of the hysteresis loop can be exactly derived. According to Ref.\cite{DSS1997}, the magnetization $m(H)$ along the lower half (ascending) branch is given by 
\be
\label{Eqmag}
\frac{1}{2}[m(H)+1]=\sum_{k=0}^{z}{z \choose k}  P^*(H)^{k}[1-P^*(H)]^{z-k}p_k(H)  
\ee
where $p_k(H)$ ($k=0..z$) is the probability for a down spin  to flip up at the field $H$ when $k$ of its $z$ nearest neighbors are up, 
\be
\label{Eqpk}
p_k(H)=\int_{(z-2k)J-H}^{+\infty}\rho(h)dh=\frac{1}{2}\mbox{erfc}\left(\frac{(z-2k)J-H}{\sqrt{2\Delta}} \right) \  ,
\ee
(here $\mbox{erfc}(x)=(2/\sqrt{\pi})\int_x^{\infty} du \exp(- u^2)$ is the complementary error function), and $P^*(H)$  is solution of the self-consistent equation
\begin{equation}
\label{Eqselfcons}
P^*(H)=\sum_{k=0}^{z-1}  {z-1 \choose k} P^*(H)^{k}[1-P^*(H)]^{z-1-k}p_k(H) \ .
\end{equation}
This key quantity represents the conditional probability that a nearest neighbor of spin $i$ flips up before spin $i$. For $z\ge 4$, the polynomial equation (\ref{Eqselfcons}) has several solutions at low enough disorder (for $\Delta<\Delta_c(z)$) and the magnetization displays a jump discontinuity at a coercive field $H_c(\Delta)$.

In the following we are interested in calculating  the correlations along the loop between the spin at site $i$ and the spin or the random field at site $j$, 
\begin{align}
G^{ss}_{ij}&=\overline{S_i}\overline{S_j} - \overline{S_i} \ \overline{S_j} \nonumber\\
G^{sh}_{ij}&=\overline{S_i h_j}
\end{align}
where the overbar denotes the average over the random field distribution $\rho(h)$ and the dependence on the applied field $H$ is implicit (hence $\overline{S_i}\equiv m(H)$ as given by Eq.(\ref{Eqmag}) along the ascending branch). Due to the average over disorder, the two functions only depend on the distance  between the two spins,  {\it i.e.} on $n$, the number of bonds between $i$ and $j$. We thus denote them by $G^{ss}(n)$ and $G^{sh}(n)$, respectively. 

Let us recall that at finite temperature and  equilibrium, because of the additional average over thermal fluctuations, there are two distinct spin-spin correlation functions, $\overline{<S_i S_j> - <S_i> <S_j>}$ and $\overline{<S_iS_j>}-\overline{<S_i>}\ \overline{<S_j>}$ where $<...>$ denotes the thermal average\cite{N1998}.  The former (the so-called connected or truncated function) may be non-zero  at $T = 0$ if the ground state of the system is highly degenerate. This does not occur when the random-field distribution is continuous and then only the disconnected function $G^{ss}(n)$ remains non-zero.   At $T=0$, one may also consider an average over all the metastable states at a given field $H$ and then distinguish again connected and disconnected contributions\cite{RT2010}.  However, the connected contribution vanishes along the hysteresis loop since there is only one metastable state and, again, only $G^{ss}(n)$ remains. On a regular Euclidian lattice, its Fourier transform is the structure factor ${\hat S}(q)$ which is the quantity  measured in scattering experiments. $G^{ss}(n)$ should not be confused with  the avalanche correlation function that measures the probability that the initial spin of an avalanche will trigger, in the same avalanche, another spin a distance $n$ away\cite{SDP2006}. In particular, in finite dimension, the algebraic decays of these two functions at criticality are not described by the same exponent.

 As was noticed only recently\cite{RT2010}, for a Gaussian distribution of the random fields, there exists  a `susceptibility sum-rule' that relates  the correlation function $G^{sh}(n)$ to the slope of the magnetization curve at $T=0$. It is obtained by using the following property of the Gaussian distribution: 
 \begin{align}
 \int dh \rho(h)h A(h)&=-\Delta \int dh \frac{d\rho(h)}{dh}A(h)\nonumber\\
& =\Delta \int dh \rho(h)\frac{\partial A(h)}{\partial h} \ .
 \end{align}
Hence
\be
\overline{S_ih_j}=\Delta\ \frac{\partial \overline{S_i}}{\partial h_j} \ ,
\ee
and by summing over $i$ and $j$ one gets
\begin{align}
\frac{1}{N}\sum_{i,j} G^{sh}_{ij}=\Delta\frac{dm}{dH}  \ .
\end{align}
On the Bethe lattice, this becomes
\begin{align}
\label{Eqsus}
 G^{sh}(0) +\sum_{n=1}^{\infty} c_nG^{sh}(n) =\Delta\frac{dm}{dH}  
\end{align}
where $c_n=z(z-1)^{n-1} $ is the number of sites distant from an arbitrary site $i$ by $n\ge 1$ bonds ({\it i.e.} the number of sites that belong to $n$th shell). 

To compute the correlation functions we first consider the case of a 1D chain ({\it i.e.} $z=2$) and then extends the results to the Bethe lattice with generic coordination number $z$. We find that 
\begin{subequations}
\begin{align}
\label{Eqresults}
G^{ss}(n)&=\lambda^{n-1}[a+b(n-1)] \\
G^{sh}(n)&=\lambda^{n-1}G^{sh}(1)
\end{align}
\end{subequations}
for $n\ge 1$  (with $G^{ss}(0)=1-m^2$ due to the hard-spin condition $S_i^2=1$, and $G^{sh}(0)$  given by Eq. (\ref{EqGsh0})). The explicit expressions of $\lambda$, $G^{ss}(1)\equiv a$, $G^{ss}(2)\equiv\lambda(a+b)$, and $G^{sh}(1)$ are given by Eqs. (\ref{Eqlambda}), (\ref{EqGss1}), (\ref{EqGss2}) and (\ref{EqGsh1}), respectively. 
 
 \subsection{One dimension}
 
\begin{figure}[hbt]
\begin{center}
\includegraphics[width=7cm]{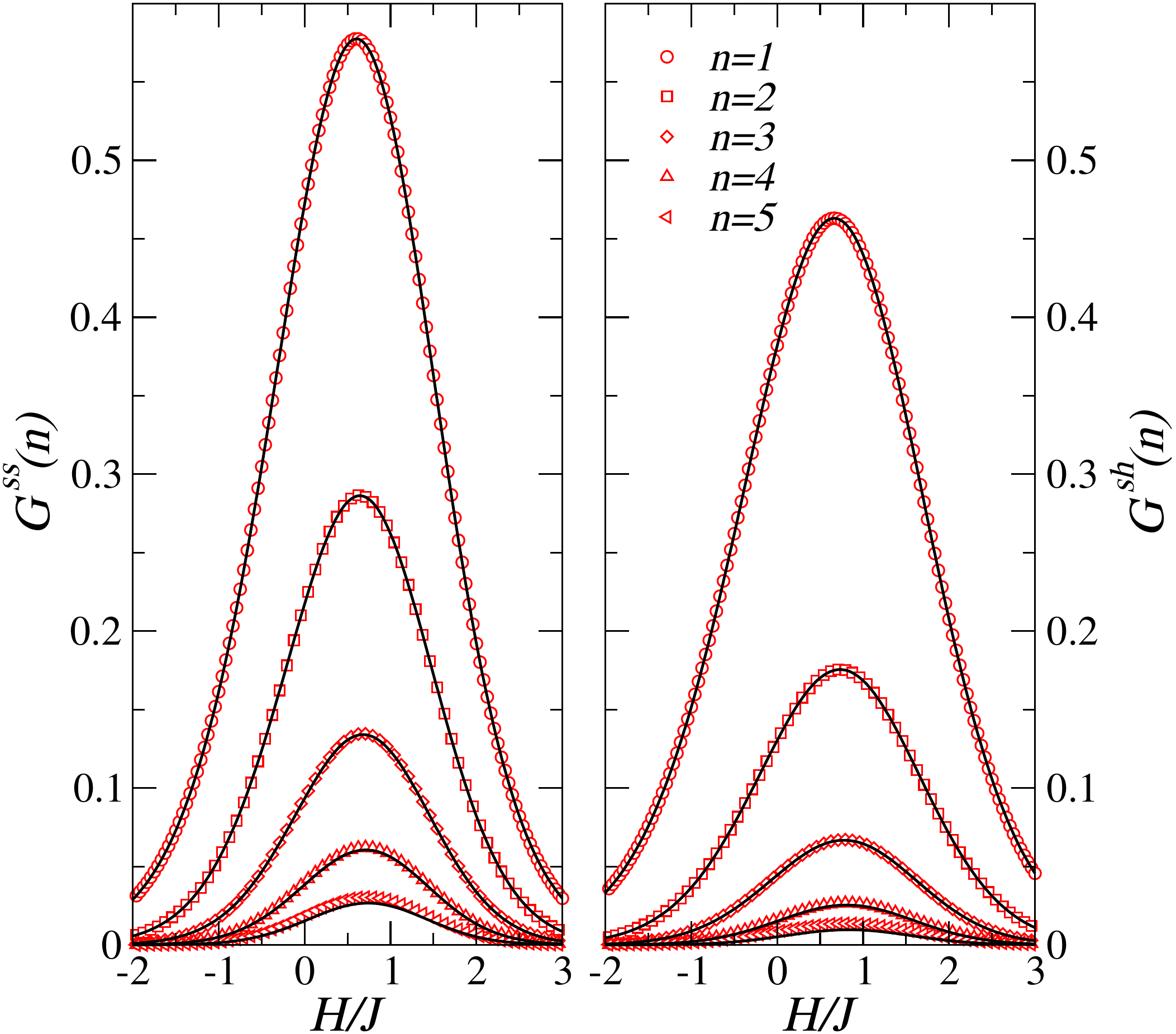}
 \caption{\label{fig1} (Color on line) Correlation functions $G^{ss}(n)$ and  $G^{sh}(n)$ along the ascending branch of the hysteresis loop for $z=2$ and $\Delta=4$.  The simulation results (symbols) are compared to the predictions of Eqs. (12) (lines). Simulations were performed on random graphs with $N=10^6$ and the results were averaged over $1000$ disorder realizations.}
\end{center}
\end{figure}

The hysteresis loop in the 1D chain was calculated in Ref.\cite{S1996}. In this case, there are only three probabilities $p_0,p_1, p_2$ defined by Eq. (\ref{Eqpk}), and from Eqs. (\ref{Eqmag}) and (\ref{Eqselfcons}) the magnetization along the ascending branch is simply given by
\begin{align}
m=2\left[(1-P^*)^2 p_0+2P^*(1-P^*)p_1+P^{*2} p_2\right]-1
\end{align}
with $P^*=p_0/(1-p_1+p_0)$ (hereafter, to simplify the notation, the dependence of all quantities on the field $H$ is dropped).  The results on the descending branch can be obtained by symmetry. The analytical calculation of the spin-spin correlation function $G^{ss}(n)$ was first considered in Ref.\cite{KFS2000} but the final expressions of $a$ and $b$ in Eq. (\ref{Eqresults}) are wrong (see Appendix A); the dependence on the distance $n$, however, is correctly given.  Moreover, $G^{sh}(n)$ was not considered. On the other hand, the exact expressions of $G^{ss}(1)=\overline{S_iS_{i+1}}-m^2$ and  $G^{sh}(0)=\overline{S_ih_i}$ were derived in Ref.\cite{ABCDDMZ2005} in order to compute the energy per spin along the hysteresis loop.
The complete calculation that leads to Eqs. (12) is performed in Appendix A. In particular, we obtain
\begin{align}
\label{Eqlambda1D}
\lambda=p_1-p_0
\end{align}
as correctly found in Ref.\cite{KFS2000}. As shown in Fig. 1,  Eqs. (12) are in excellent agreement with the results of numerical simulations performed on random graphs with $z=2$ (we also checked numerically that the expression of $G^{ss}(n)$ given in Ref.\cite{KFS2000} is not valid).

The two functions $G^{ss}(n)$ and  $G^{sh}(n)$ are  thus characterized by the same correlation length $\xi_{\lambda}=(-\ln \lambda)^{-1}=[-\ln(p_1-p_0)]^{-1}$. However,  $G^{ss}(n)$ is not a purely exponential function because of the prefactor $b(n-1)$. Remarkably, a similar behavior has been observed for the {\it equilibrium} RFIM in the very few cases where the correlation function $G^{ss}_{eq}(n)=\overline{<S_iS_{i\pm n}>}-\overline{<S_i>}\ \overline{<S_{i\pm n}>}$ has been calculated exactly. This is indeed the leading long-distance behavior observed at $T>0$ with the special random-field distribution (somewhat related to percolation) considered in Ref.\cite{GM1983} (in this model, however, the $T=0$ behavior is  complicated and the correlation function behaves  at long distance as an exponential divided by $n^2$\cite{LN1989}).  For the Gaussian distribution that we here consider, no analytical expression is available for generic values of $\Delta$ and $T$, but an exact result has been obtained in the universal regime where the random field and the temperature are both much smaller than the exchange coupling\cite{FLM2001}. For $H=0$, the leading long-distance behavior turns out to be also proportional to $n\exp{(-n/\xi_{eq})}$, where $\xi_{eq}=8J^2/(\pi^2 \Delta)=(2/\pi^2) L_{IM}$ and $L_{IM}$ is the Imry-Ma length that sets the  typical size of the domains in the 1D chain at $T=0$\cite{IM1975}. This coincidence is noteworthy but it must emphasized that the full expression of $G^{ss}_{eq}(n)$ in this  regime is  much more complicated than the one described by Eq. (\ref{Eqresults}) (moreover, for $H\ne 0$, $G^{ss}_{eq}(n)$ decays as a sum of exponentials).  The correlation length $\xi_{\lambda}$ along the hysteresis loop  also behaves quite differently from $\xi_{eq}$ in the  limit $\Delta \ll J $: it goes to the finite value $1/\ln(2)$ in zero applied field (as $H=0$ does not play any special role along the hysteresis loop) and grows like $\xi_{\lambda}\sim  (\sqrt{\pi}J/\sqrt{2\Delta})\exp(J^2/2\Delta)$ for $H=J$, which is the value of the  field for which the susceptibility $\partial m/\partial H$ is maximum. 

An interesting consequence of Eq. (\ref{Eqresults}) is that the structure factor ${\hat S}(q)$ in the small-$q$ regime is a superposition of a Lorentzian and a Lorentzian-squared terms. By definition
\begin{align}
{\hat S}(q)=G^{ss}(0)+ \sum_{n=1}^{\infty}[ e^{iqn}+ e^{-iqn}]G^{ss}(n) \ ,
\end{align}
and using
\begin{align}
\sum_{-\infty}^{+\infty} e^{iql} \lambda^{\vert l \vert }=\frac{\sqrt{1-x^2}}{1-x\cos q}
\end{align}
with
\begin{align}
\label{Eqx}
x=\frac{2\lambda}{1+\lambda^2}\ ,
\end{align}
we  obtain after simple algebra
\begin{align}
\label{EqSq}
{\hat S}(q)=A+\frac{B}{1-x\cos q}+\frac{C}{[1-x\cos q]^2}
\end{align}
with
\begin{align}
A&=1-m^2+\frac{b-a}{\lambda}\nonumber\\
B&=\frac{(a-b)\sqrt{1-x^2}-b}{\lambda}\nonumber\\
C&=b\frac{1-x^2}{\lambda} \ .
\end{align}

The Lorentzian plus Lorentzian-squared structure that emerges from Eq. (\ref{EqSq}) in the small-$q$ regime is also found in the mean-field theory of the equilibrium RFIM\cite{KW1981} and is usually used to fit experimental data on random magnetic systems.  

In contrast, the spin-random field correlation function $G^{sh}(n)$ is a pure exponential  for $n\ge 1$ so that its Fourier transform simply reads
\begin{align}
\label{EqGshq}
\hat{G}^{sh}(q)=[G^{sh}(0)-\frac{G^{sh}(1)}{\lambda}]+\frac{G^{sh}(1)}{\lambda}\frac{\sqrt{1-x^2}}{1-x\cos q} \ .
\end{align}
This  yields
\begin{align}
\hat{G}^{sh}(q=0)= G^{sh}(0)+\frac{2}{1-\lambda}G^{sh}(1) \ ,
\end{align}
and using  the expression of the magnetization, Eq. (\ref{Eqmag}), and Eqs. (\ref{EqGsh0}) and (\ref{EqGsh1}) for $G^{sh}(0)$ and $G^{sh}(1)$, one can  check that the susceptibility sum-rule, $\hat{G}^{sh}(q=0)=\Delta (dm/dH)$, is indeed satisfied. Note  that the $q$-independent term inside brackets in Eq. (\ref{EqGshq}) is non-zero, which is a somewhat unusual feature (the constant term $A$ in Eq. (\ref{EqSq})  is also non-zero because Eq. (\ref{Eqresults}) is only valid for $n\ge 1$). As will be discussed in more detail in section III in the case of the Bethe lattice, this has  a significant consequence for the matrix inverse of $G^{sh}$ (the so-called direct correlation function).

\subsection{Bethe lattice}

\begin{figure}[hbt]
\begin{center}
\includegraphics[width=7cm]{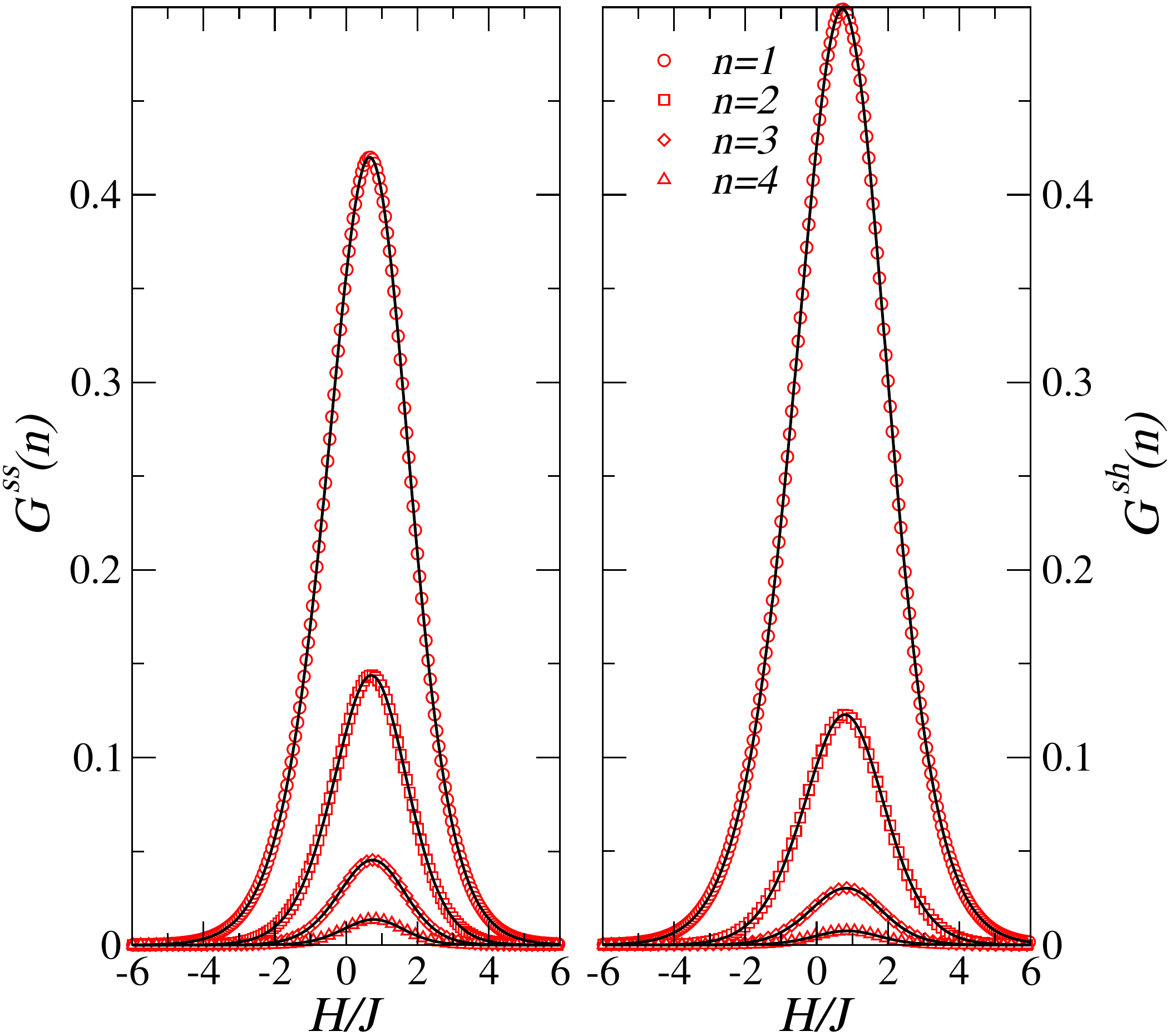}
 \caption{\label{fig2} (Color on line ) Correlation functions $G^{ss}(n)$ and  $G^{sh}(n)$  on a Bethe lattice with coordination number $z=3$ for $\Delta=9$ (the curves result from an average over $5000$ random graphs of size $N=10^5$).  The simulation results (symbols) are compared to the predictions of Eqs. (12) (lines).}
\end{center}
\end{figure}
\begin{figure}[hbt]
\begin{center}
\includegraphics[width=7cm]{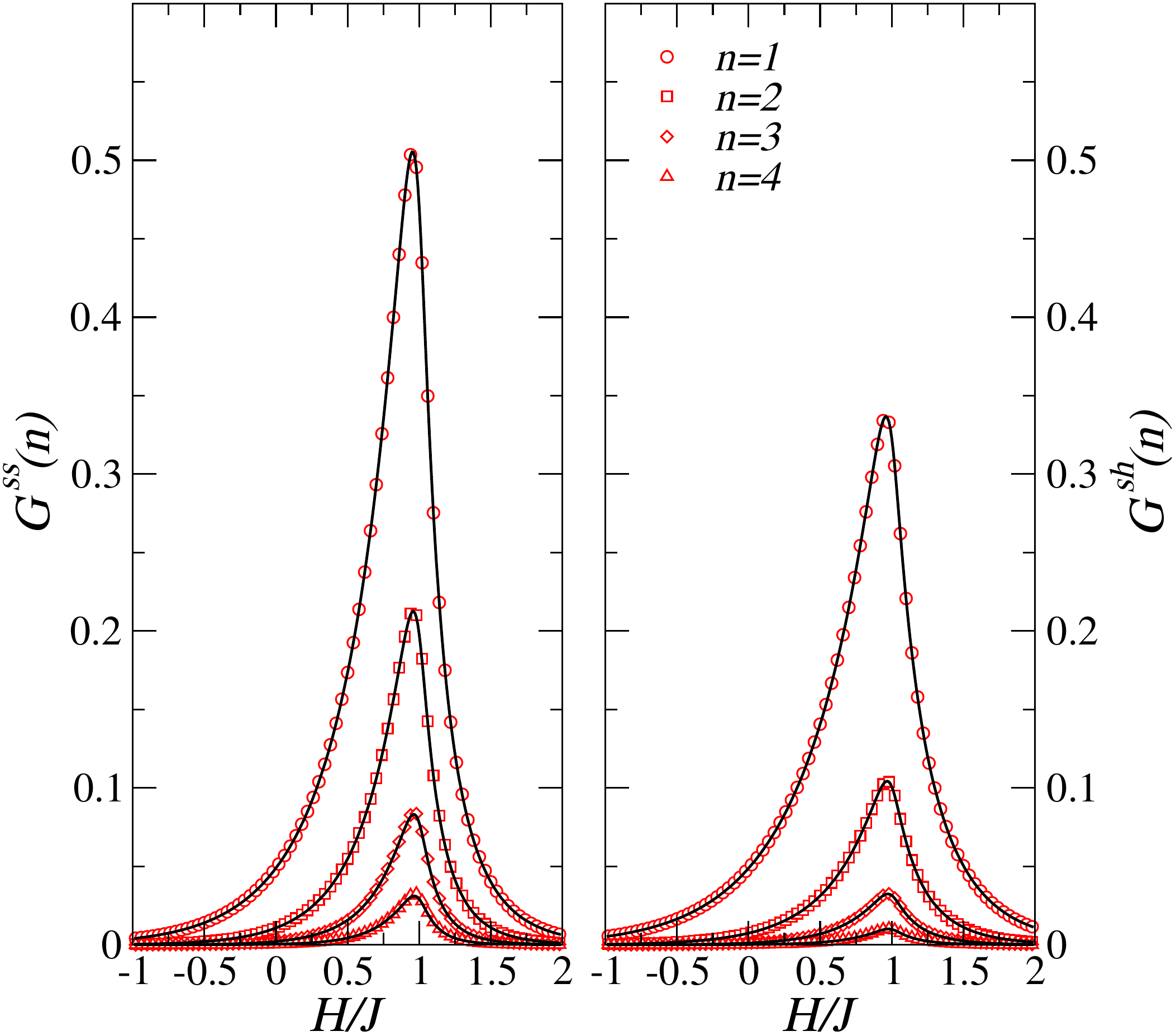}
 \caption{\label{fig3}(Color on line) Same as Fig. \ref{fig2} for  $z=4$ and $\Delta=4$.}
\end{center}
\end{figure}

In principle, the  probabilistic reasoning used in Appendix A for the one-dimensional chain can be extended to the case of the Bethe lattice with generic coordination number $z$.  This is how the analytical expressions of $G^{ss}(1)$ and $G^{sh}(0)$ were derived in Ref.\cite{IOV2005} in order to compute the energy per spin along the hysteresis loop (thereby generalizing the 1D results of Ref.\cite{ABCDDMZ2005}). These expressions are recalled in Appendix B where we also calculate $G^{ss}(2)$ and $G^{sh}(1)$. However, using the same method to derive the general expressions of $G^{ss}(n)$ and $G^{sh}(n)$ is unnecessarily complicated. Instead, one can simply exploit the fact that there is a unique path connecting a given pair of spins on a Bethe lattice so that the dependence of the correlation functions on the distance $n$ must be the same as in one-dimension (just like in nonrandom systems). This implies that Eqs. (12) are also valid for the Bethe lattice.  Strictly speaking, we do not provide a demonstration of this assertion\footnote[1]{Eq. (\ref{Eqresults})  for $G^{ss}(n)$ is  actually confirmed by the very  recent analytical calculations of Ref.\cite{HPT2011}. In that work, however, there is some confusion between $G^{ss}(n)$ and the so-called `avalanche correlation function' (using the terminology of Ref.\cite{SDP2006}). This latter function can be shown to behave as a simple exponential for $n\ge 1$, without the $n-1$ prefactor.} but it is fully supported by numerical simulations for small values of $n$, as illustrated in Figs. 2 and 3. 

The analytical expression of $\lambda$ can then be obtained via the susceptibility sum-rule. Inserting Eq. (12b) in Eq. (\ref{Eqsus}) yields
\begin{align}
\label{Eqsus1}
\Delta\frac{dm}{dH}=G^{sh}(0) +\frac{z}{1-(z-1)\lambda}G^{sh}(1) 
\end{align}
so that\footnote[2]{Using  Eqs. (\ref{Eqmag}), (\ref{EqGsh0}), and (\ref{EqGsh1}), it can be checked that Eq. (\ref {Eqlambda}) is equivalent to the compact expression obtained in Ref.\cite{HPT2011}: $\lambda=(z-1)^{-1} \partial F(P^*)/\partial P^*$, where $F(P^*)$ is the  the r.h.s. of Eq. (\ref{Eqselfcons}).}
\begin{align}
\label{Eqlambda}
\lambda=\frac{1}{z-1}[1-z\frac{G^{sh}(1)}{\Delta \: dm/dH-G^{sh}(0)}]
\end{align}
Using Eqs. (\ref{Eqmag}), (\ref{EqGsh0}) and (\ref{EqGsh1}), one can  check that Eq. (\ref{Eqlambda1D}) is recovered for $z=2$. 
\begin{figure}[hbt]
\begin{center}
\includegraphics[width=8cm]{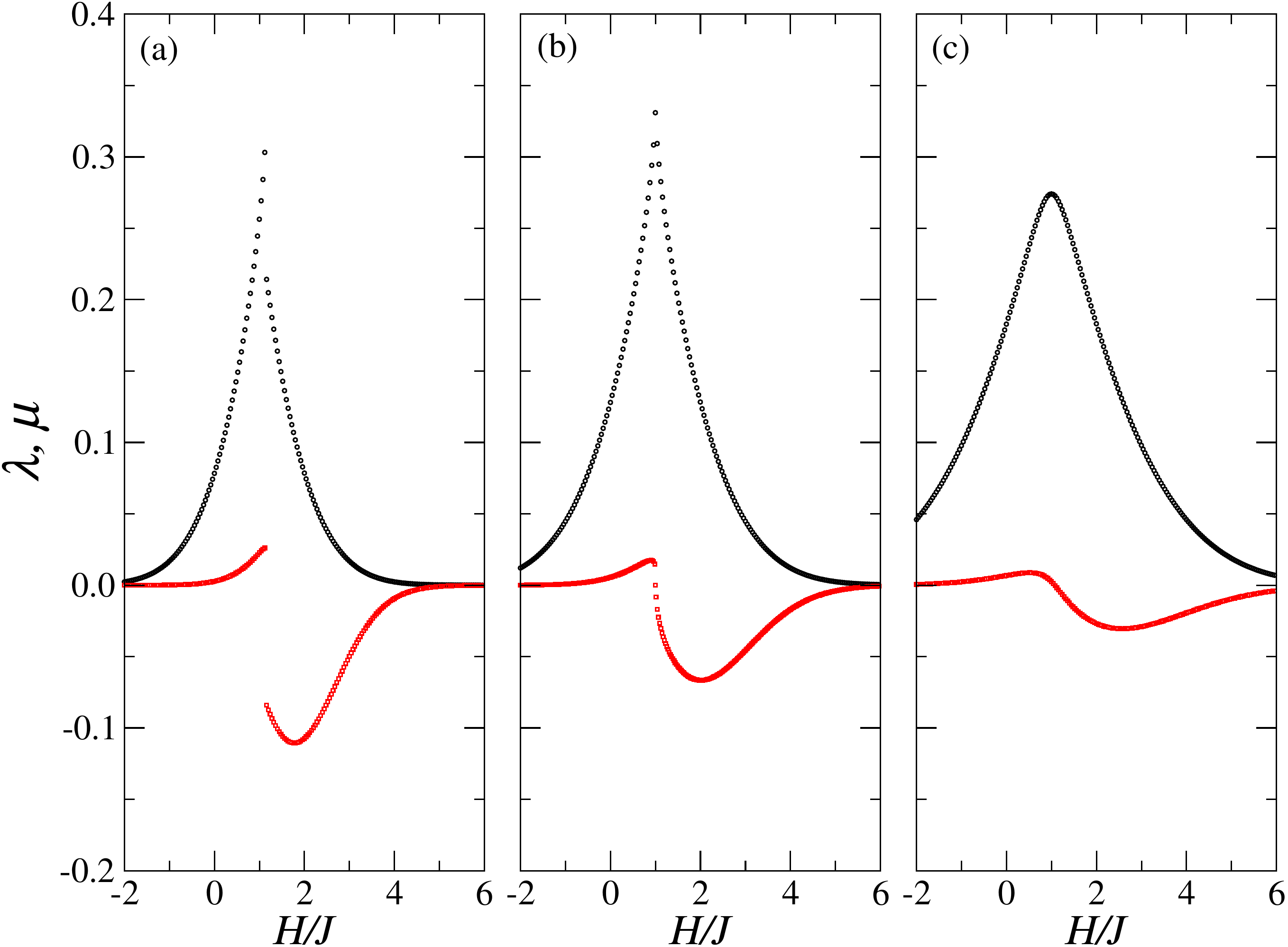}
 \caption{\label{fig4} (Color on line) The quantities $\lambda(H)$ (top) and $\mu(H)$ (bottom) characterizing the exponential decay of the correlation and direct correlation functions, respectively,  for $z=4$ and  (a) $\Delta=2$, (b) $\Delta=\Delta_c\approx 3.173$, and (c) $\Delta=6$. In (a), below the critical disorder $\Delta_c$, $\lambda$  and $\mu$ jump discontinuously at the coercive field $H_c(\Delta)$.  In (b), $\lambda\rightarrow (z-1)^{-1}=1/3$ at the critical field $H_c(\Delta_c)=J$. Note  that $\mu$ is significantly smaller than $\lambda$ in (c), above the critical disorder. }
 \end{center}
\end{figure}

For $z\ge 4$ and $\Delta<\Delta_c(z)$, the magnetization jumps discontinuously at the coercive field $H_c(\Delta)$, which corresponds to a spinodal singularity where $dm/dH \rightarrow +\infty$\cite{DSS1997,CGZ2002}. As can be deduced from Eq. (\ref{Eqsus1}),  this is due to the fact that $\lambda \rightarrow (z-1)^{-1}$, as illustrated in Fig. 4b for $z=4$. Therefore the correlation length $\xi_{\lambda}=(-\ln \lambda)^{-1}$ keeps a finite value at all $H$ and $\Delta$, including at the critical point. This is indeed the expected (and standard) behavior on the Bethe lattice where the divergence of the susceptibility is generated by the exponential growth of the number of sites at the distance $n$ (due to the hyperbolic-like geometry of the lattice) and is not associated to a divergence of the correlation length. 

\section{Direct correlation functions}

We now investigate the structure of the direct correlation functions (or one-particle irreducible functions, or else proper vertices in field-theoretic language) which (roughly speaking) are the matrix inverses of the correlation functions $G^{sh}_{ij}$ and $G^{sh}_{ij}$ (see Eqs. (\ref{OZ1}) and (\ref{OZ2}) below). As briefly mentioned in the introduction, the motivation for this calculation is that  proper vertices may be simpler or at least shorter-ranged than the Green's functions, and therefore can be used as the building blocks of approximate theories. For instance, in liquid-state theory, the direct correlation function $c(r)$, which is the matrix inverse of the pair correlation function $h(r)$ at equilibrium and is defined via the so-called Ornstein-Zernike (OZ) equation, is in general shorter-ranged than $h(r)$ (essentially having the range of the pair potential), irrespective of the thermodynamic state\cite{HM2006}. This feature is the starting point of the successful integral equation approach to the structure and thermodynamics of simple liquids. For Ising spins on a lattice with nearest-neighbor interactions, it is also a reasonable approximation to assume that  the matrix inverse of the spin-spin correlation function is zero for $n>1$\cite{note1}. This can been used to build a very accurate description of the  three-dimensional Ising model\cite{HS1977} and other spin models\cite{GKRT2001}, including in the presence of quenched disorder\cite{KRT1997}. More recently, a similar approximation has  been proposed to obtain an analytical description of the hysteresis loop in the three-dimensional soft-spin random field model at $T=0$ \cite{RT2010}. 
It is therefore interesting to check whether the direct correlation functions on the Bethe lattice are indeed shorter-ranged than $G^{sh}$ and $G^{sh}$.

We first consider the function ${\bf C}^{sh}=\{C^{sh}_{ij}\}$  defined by  the OZ equation $\sum_k C^{sh}_{ik}G^{sh}_{kj}=\delta_{ij}$, {\it i.e.}
 \begin{align}
\label{OZ1}
 {\bf C}^{sh}=[{\bf G}^{sh}]^{-1} 
 \end{align}
where ${\bf G}^{sh}$ and ${\bf C}^{sh}$ are $N\times N$ matrices. 
As  pointed out in the preceding section,  $G^{sh}(n)$ is a pure exponential for $n>1$, but $G^{sh}(1)\ne \lambda G^{sh}(0)$. At first sight this is an innocuous feature but it has an important consequence for $C^{sh}(n)$. Indeed, as is well known (and is also shown below), if $G^{sh}(1)$ were equal to $\lambda G^{sh}(0)$ and therefore $G^{sh}(n)=\lambda^n G^{sh}(0)$ for all $n$,  ${\bf C}^{sh}$ would be  simply proportional to  ${\bf A}$, the adjacency matrix of the lattice ($A_{ij}=1$ is the vertices $i$ and $j$ are connected and $0$ otherwise) and the range of $C^{sh}(n)$ would then be limited to the n.n. distance.

In order to solve the `Ornstein-Zernike' equation (\ref{OZ1}), it is convenient to consider the simple random walk on the lattice where, at each time step, a particle jumps to any of the $z$ neighboring sites with probability $1/z$. Indeed, ${\bf G}^{sh}$ is  directly related to the lattice Green function ${\bf F}(x)$ which is the probability generating function  defined as (see {\it e.g.} Refs.\cite{G1994,MT1996})
\begin{align}
F_{ij}(x)=\sum_{\tau=0}^{\infty}x^{\tau}p^{\tau}_{ij}
\end{align}
where $p^{\tau}_{ij}$ is the probability that the particle starting at $i$ reaches $j$ after $\tau$ time steps.   As is well known, one has 
\begin{align}
{\bf F}(x)=({\bf I}-x{\bf M})^{-1}
\end{align}
 where ${\bf M} =(1/z){\bf A}$. Since all sites are topologically equivalent, one can choose the origin of coordinates as the origin of the random walk and simply consider $F_n(x)=\sum_{\tau=0}^{\infty}x^{\tau}f_{\tau}(n)$ where $f_{\tau}(n)$ is the probability of being in the $n$th shell after $\tau$ time steps. By definition, $F_0(x)=F_{ii}(x)$ and $F_n(x)=c_n F_{ij}(x)$ where $i$ and $j$ are connected by $n\ge 1$ bonds (recall that $c_n=z(z-1)^{n-1}$).
 It can then be shown\cite{MT1996} that
\begin{align}
F_0(x)&=\frac{2(z-1)}{z-2+\sqrt{z^2-4(z-1)x^2}}\nonumber\\
F_n(x)&=c_n\Big(\frac{r(x)}{z-1}\Big)^{n}\ F_0(x)\ \ \ \mbox{for} \ n\ge 1 
\end{align}
with
\begin{align}
\label{Eqrx}
r(x)=\frac{z-\sqrt{z^2-4(z-1)x^2}}{2x}   \ .
\end{align}
Using the identification 
\begin{align}
\label{lambdaBethe}
\lambda=\frac{r(x)}{z-1} \ , 
\end{align}
we  readily see from Eq. (12b) that
\begin{align}
{\bf G}^{sh}=u{\bf F}(x)+v{\bf I}
\end{align}
with
\begin{align}
u&= \frac{G^{sh}(1)}{\lambda F_0(x)}\nonumber\\
v&= G^{sh}(0) -\frac{G^{sh}(1)}{\lambda} .
\end{align} 
Note that Eq. (\ref{lambdaBethe}) can  also be written as 
\begin{align}
\lambda=\frac{F_1(x)}{xF_0(x)}=\frac{F_0(x)-1}{xF_0(x)} 
\end{align}
which can be inverted to express $x$ as a function of $\lambda$,
 \begin{align}
x=\frac{z\lambda}{1+(z-1)\lambda^2}  \ .
\end{align}
Therefore $x\rightarrow 1$ when $\lambda \rightarrow (z-1)^{-1}$ at the spinodal.

If $v$ were equal to $0$,  one would  simply have ${\bf C}^{sh}=[u{\bf F}]^{-1}=u^{-1}({\bf I}-x{\bf M})$ and $C^{sh}(n)$ would be zero for $n>1$, as stressed above. The matrix equation ${\bf C}^{sh}=[u{\bf F}+v{\bf I}]^{-1}$ is  now easily solved:
 \begin{align}
{\bf C}^{sh}&=\frac{1}{u}[{\bf I}+\frac{v}{u}{\bf F}^{-1}(x) ]^{-1}{\bf F}^{-1}(x) \nonumber\\
&=\frac{1}{u}[{\bf I}+\frac{v}{u}({\bf I}-x{\bf M})]^{-1}({\bf I}-x{\bf M})\nonumber\\
&=\frac{1}{u+v}[{\bf I}-\frac{vx}{u+v}{\bf M})]^{-1}({\bf I}-x{\bf M})\nonumber\\
&=\frac{1}{u+v}{\bf F}(x')({\bf I}-x{\bf M})
\end{align}
where
\begin{align}
x'=\frac{v}{u+v}x \ .
\end{align} 
This yields
\begin{align}
C^{sh}_{ij}=\frac{1}{u+v}[F_{ij}(x')-\frac{x}{z}\sum_{k/j}F_{ik}(x')] 
\end{align} 
where $k$ is connected to $j$. Hence
\begin{align}
C^{sh}(0)&=\frac{1}{u+v}[F_0(x')-\frac{x}{z}F_1(x')] =-\frac{u}{v(u+v)}F_0(x')+\frac{1}{v}\nonumber\\
C^{sh}(1)&=\frac{1}{u+v}[F_1(x')-\frac{x}{z}(F_0(x')+\frac{F_2(x')}{z})]\nonumber\\
&=-\frac{u}{v(u+v)}\frac{F_0(x')-1}{x'} 
\end{align} 
and
\begin{align}
C^{sh}(n)&=\frac{1}{u+v}\Big[\frac{F_n(x')}{c_n}-\frac{x}{z}[(z-1)\frac{F_{n+1}(x')}{c_{n+1}}+\frac{F_{n-1}(x')}{c_{n-1}}]\Big]\nonumber\\
&=\frac{1}{u+v}\frac{1}{c_n}\Big[F_n(x')-\frac{x}{z}[F_{n+1}(x')+(z-1)F_{n-1}(x')]\Big]\nonumber\\
&=-\frac{u}{v(u+v)}\frac{F_n(x')}{c_n} \ \ \mbox{for}\  \ n\ge 2
\end{align} 
where we have used the recurrence relations\cite{MT1996}
\begin{align}
F_2(x')&=\frac{z}{x'}F_{1}(x')-zF_{0}(x')\nonumber\\
F_{n+1}(x')&=\frac{z}{x'}F_{n}(x')-(z-1)F_{n-1}(x') \ \ \mbox{for} \ n\ge 2 \ .
\end{align} 
Introducing
\begin{align}
\label{Eqmu}
\mu=\frac{F_0(x')-1 }{x'F_0(x')} 
\end{align} 
which is equivalent to
\begin{align}
x'=\frac{z\mu }{1+(z-1)\mu^2} \ \ ,
\end{align} 
we finally obtain
\begin{align}
C^{sh}(0)=\frac{1}{u+v}F_0(x')(1-\mu x)
\end{align} 
and
\begin{align}
\label{EqCsh}
C^{sh}(n)=-\frac{u}{v(u+v)}F_0(x')\mu^{n} \ \ \mbox{for} \ n\ge 1 \ .
\end{align} 
For $z=2$, one has $F_0(x)=(1-x^2)^{-1/2}$ and one can check that Eq. (\ref{EqCsh}) is in agreement with the expression obtained  by directly solving the OZ equation in Fourier space. One can also check that  the following susceptibility sum-rule is satisfied,
\begin{align}
\Delta\frac{dm}{dH}=[C^{sh}(0) +\frac{z}{1-(z-1)\mu}C^{sh}(1)]^{-1} \ .
\end{align} 

We thus see from Eq. (\ref{EqCsh}) that $C^{sh}(n)$  exhibits an exponential decay like $G^{sh}(n)$,  but with a different correlation length $\xi_{\mu}= (-\ln\vert \mu\vert)^{-1}$. Moreover, since  the sign of $v=G^{sh}(0) -G^{sh}(1)/\lambda$ depends on $H$, $\mu$  is not always positive (see Fig. 4) and there is a range of $H$ where the exponential decay is modulated by an oscillating sign. It turns out, however, that $\mu$ is significantly smaller than $\lambda$ in the weak-coupling (or large-disorder) regime, as illustrated in Fig. 4c, so that $C^{sh}(n)$  decreases  rapidly with $n$.  Indeed, expanding all quantities in powers of $J$, one finds that
\begin{align}
G^{sh}(1)=\lambda G^{sh}(0)+O(J^3)
\end{align}
so that $\mu=O(J^3)$. Since $C^{sh}(1)=-J/\Delta+O(J^2)$, this implies that $C^{sh}(2)=O(J^4)$, $C^{sh}(3)=O(J^7)$, etc... Therefore, setting $C^{sh}(n)= 0$ for $n>1$ may be a reasonable approximation above the critical disorder (for instance, one has $J/\Delta_c\approx 0.315$ for $z=4$ at the critical disorder).

A similar calculation can be performed for the direct correlation function $C^{ss}(n)$ associated to the spin-spin correlation function $G^{ss}(n)$. It is defined via a second OZ equation 
\begin{align}
\label{OZ2}
{\bf C}^{ss}=-{\bf C}^{sh}{\bf G}^{ss}{\bf C}^{sh} 
\end{align} 
whose origin (in terms of a  Legendre transform) is explained in Ref. \cite{RT2010}. After some involved algebra, we obtain
\begin{align}
C^{ss}(n)=\mu^{n-1}[a'+b'(n-1)]
\end{align}
for $n\ge 1$, were $a',b'$ are functions of $H/J$ and $\Delta/J$ which are not detailed here for the sake of brevity. Hence $C^{ss}(n)$ has the same structure as $G^{ss}(n)$ with $\lambda$ replaced by $\mu$ (the fact that there is only one correlation length appearing in the final result and not two as could be expected from Eq. (\ref{OZ2}) is due to some remarkable cancellations occurring in the intermediate steps of the calculation). As a consequence,  $C^{ss}(n)$ decreases rapidly with $n$ like $C^{sh}(n)$ ({\it i.e.} $C^{ss}(2)=O(J^4)$, $C^{ss}(3)=O(J^7)$, etc...), so that the n.n. approximation is also reasonable above $\Delta_c$.

\section{Summary and conclusion}

In this work we have determined the two-point spin-spin and spin-random field correlation (or Green's) functions of the zero-temperature Gaussian RFIM  on a Bethe lattice along the saturation hysteresis loop. This adds the model to the short list of nonequilibrium systems for which  the correlation functions are analytically calculable. In the RFIM, these functions are not  known at equilibrium, even in one dimension, except for very special random-field distributions or in the universal regime of very small disorder. We find that the two correlation functions decay exponentially with the distance with  the same correlation length. This length remains finite at the disorder-induced critical point, which is the expected behavior on a Bethe lattice. The spin-spin correlation function also contains a prefactor proportional to the distance, so that the corresponding structure factor is a sum of a Lorentzian and a Lorentzian squared  at small wavevector,  just like in the mean-field description of the equilibrium RFIM. This gives some justification for using these simple functional forms to describe the data obtained in scattering experiments in RFIM-like systems ({\it e.g.} along the adsorption-desorption isotherms in the case of gases  adsorbed in disordered porous solids). We also find  that the direct correlation functions, which are the inverses of the correlation functions in the sense of matrices, have essentially the same analytic structure as the correlation functions, but with a different correlation length and a modulation of the sign (depending on the value of the applied field). This correlation length, however, is small, especially in the weak-coupling/large-disorder regime, and it is thus reasonable to assume that the range of the direct correlation functions  is limited to the nearest-neighbor distance above the critical disorder. Although this `Ornstein-Zernike'  type of approximation breaks down in the vicinity of the critical point\cite{note2}, it may the starting point of an analytical description of the hysteresis loop in the three-dimensional RFIM, as  developed recently for the soft-spin version of the model\cite{RT2010}.

\appendix
\section{Calculation of $G^{ss}(n)$ and $G^{sh}(n)$ in the one-dimensional chain}

In this appendix, we present the calculation of the correlation functions $G^{ss}(n)$ and $G^{sh}(n)$ 
along the hysteresis loop in the 1D chain (specifically, along the ascending branch  obtained by starting 
with a field $H$ large and negative). The correlations are obtained by generalizing 
the procedure used in Ref.\cite{S1996} to get the magnetization.

We first consider the spin-spin correlation function $G^{ss}(n)=\overline{S_0S_n}-\overline{S_0}^2$. 
As stressed in the main text, the calculation of $G^{ss}(n)$ was first considered in Ref.\cite{KFS2000} but
the final expression is flawed. We therefore redo the whole calculation, closely following the reasoning and the notations of Ref.\cite{KFS2000} (note however that the calculation in Ref.\cite{KFS2000} is performed along the descending branch of the loop). By definition, 
\begin{align}
\label{EqS0Sn}
\overline{S_0 S_n}&=  \sum_{S_0,S_n} S_0 S_n \Phi_n(S_0,S_n)\nonumber\\
&=\Phi_n(+,+)-\Phi_n(+,-)-\Phi_n(-,+)+\Phi_n(-,-) \ ,
\end{align}
where $\Phi_n(+,+)$ is the probability that spins at $0$ and $n$ are both up,
and  $\Phi_n(+,-),\Phi_n(-,+),\Phi_n(-,-)$ are defined analogously. 

To calculate the probabilities $\Phi_n(S_0,S_n)$ we relax the spins in two steps (a spin is relaxed when it is
aligned with its local field). In the first step, the spins $S_0$ and $S_n$ are kept down and the other 
spins  can relax. In the second step, we also allow $S_0$ and $S_n$ to relax. The crucial point is that 
the final state does not depend on the order in which spins are relaxed.

\begin{figure}[hbt]
\begin{center}
\includegraphics[width=8cm,clip]{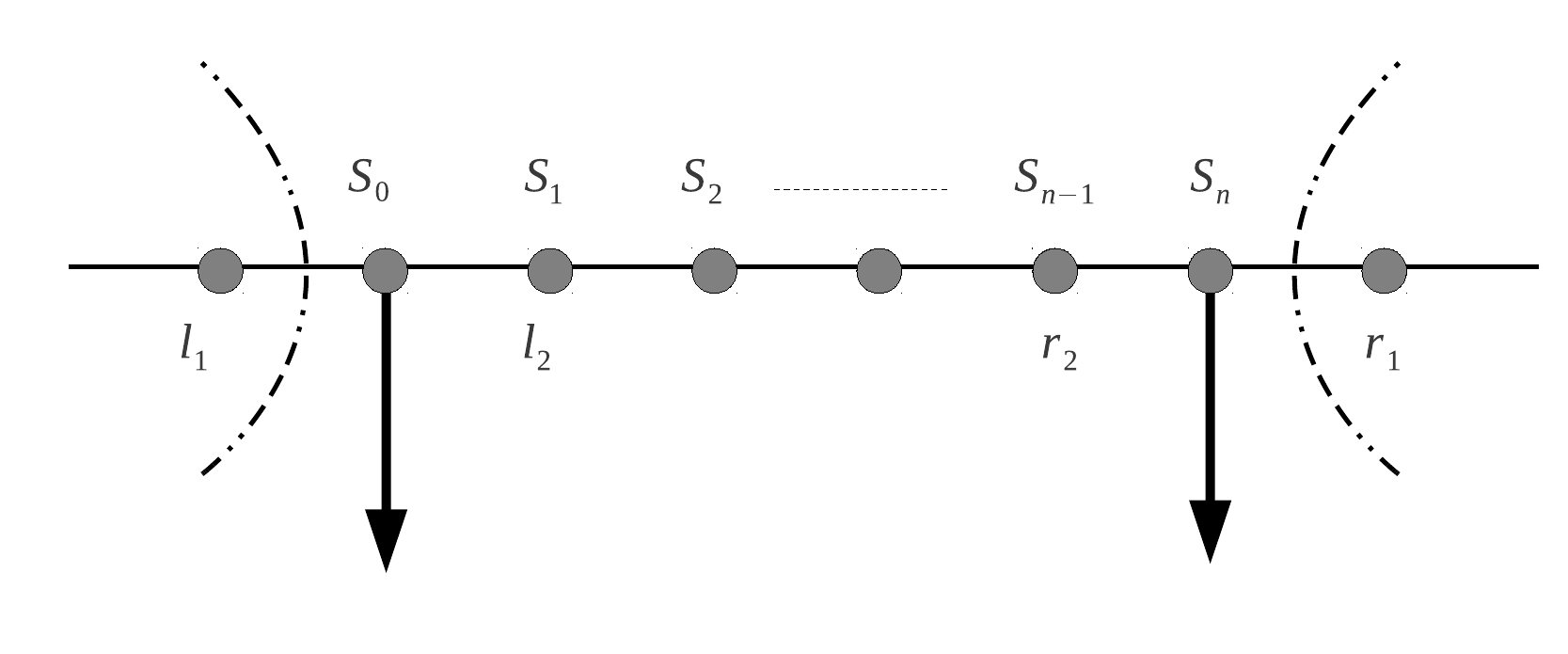}
 \caption{\label{fig5} Schematic representation of the environment of the spins $S_0$ and $S_n$. In addition to the spin variables $S_i=\pm 1$, we  also use the  variables $l_1=(1+S_{-1})/2$, $l_2=(1+S_{1})/2$, $r_2=(1+S_{n-1})/2$,  and $r_1=(1+S_{n+1})/2$ taking the values $0,1$.}
\end{center}
\end{figure}

In the first step, we need to  compute the constrained probabilities $G_n(S_1,S_{n-1})$ (not to be confused 
with the correlations functions) that the spins adjacent to $S_0$ and $S_{n}$ (see Fig. \ref{fig5} for a schematic representation) are in the state $\{S_1,S_{n-1}\}$.

\subsection{Constrained probabilities $G_n(S_1,S_{n-1})$}

By definition,
\begin{equation}
G_n(S_1,S_{n-1})=\sum_{S_2 \ldots S_{n-2}} P( S_1,S_2 \ldots S_{n-2},S_{n-1} )
\label{def_G}
\end{equation}
where $P(S_1,S_2 \ldots S_{n-2},S_{n-1})$ is the probability of the configuration $\{S_1,S_2 \ldots S_{n-2},S_{n-1}\}$ when the spins $S_0$ and $S_n$ are pinned down and the  spins between them are allowed to relax. 
 
Carrying out the calculation as in Ref.\cite{KFS2000}, one easily derives the following recurrence 
relations for the constrained probabilities along the ascending branch of the hysteresis loop,
\begin{align}
G_n(-,-)&=(1-p_0)G_{n-1}(-,-)+(1-p_1)G_{n-1}(-,+)  \nonumber \\
G_n(+,-)&=(1-p_0)G_{n-1}(+,-)+(1-p_1)G_{n-1}(+,+) \ .
\end{align}
Moreover $G_n(-,+)=G_n(+,-)$ by symmetry, and $G_n(+,+)$ is obtained via the sum-rule
\begin{equation}
G_n(-,-)+G_n(+,-)+G_n(-,+)+G_n(+,+) = 1 \ .
\end{equation}
This gives the matrix equation 
\be
{\bf G}_n={M}{\bf G}_{n-1}
\ee
where
\begin{align}
{\bf G}_n=\left[
\begin{array} {c}
G_n(-,-) \\
G_n(+,-) \\
G_n(+,+) \\
\end{array}\right]
\end{align}
and
\begin{align}
M=\left[
\begin{array} {lcr}
1-p_0   & 1-p_1     &          0 \\
   0      & 1-p_0            &    1-p_1 \\
p_0 & 2p_0+p_1-1 & 2p_1-1  \\
\end{array}
\right] \ .
\end{align}
Hence ${\bf G}_n={M}^{n-1}{\bf G}_1$,
with ${\bf G}_1 \equiv\left[\begin{array} {c}1 \\0 \\0 \\\end{array}\right]$.  
We then change to the vector basis\cite{KFS2000} 
\begin{equation}
V=
\left[
\begin{array} {ccc}
(1-P^*)^2   & +1 & +1 \\
P^*(1-P^*)     & -1 &  0 \\
  P^{*2}          & +1 & -1  \\
\end{array}
\right]
\end{equation}
(recall that $P^*=p_0/(1-p_1+p_0)$) in which the matrix $M$  takes the form
\begin{equation}
\tilde{M}=V^{-1} M V 
= 
\left[
\begin{array} {ccc}
   1   &     0     &       0  \\
   0   & p_1-p_0 &  1  -p_1 \\
   0   &     0     &  p_1-p_0  \\
\end{array}
\right] 
\end{equation}
so that
\begin{equation}
\tilde{M}^{n-1}= 
\left[
\begin{array} {ccc}
   1   &     0     &       0  \\
   0   & (p_1-p_0)^{n-1} &  (n-1)(1-p_1)(p_1  -p_0)^{n-2} \\
   0   &     0     &  (p_1-p_0)^{n-1}  \\
\end{array}
\right] \ .
\end{equation}
To simplify the notation, we  shift to the variables $l_2 \equiv (1+S_1)/2$ and $r_2 \equiv (1+S_{n-1})/2$ that take the values $0,1$, and after some algebra we finally obtain

\begin{align}
G_n(0,0)&= (1-P^*)^2+ \big[ P^* (1-P^*) \nonumber\\
  &+ (n-1) P^*\frac{1-p_1}{p_1-p_0} +  P^* \big ]  (p_1-p_0)^{n-1} \nonumber \\
G_n(1,0)&= P^* (1-P^*)- \big[ P^* (1-P^*)\nonumber\\
& + (n-1) P^*\frac{1-p_1}{p_1-p_0}  \big ]  (p_1-p_0)^{n-1} \nonumber \\ 
G_n(1,1)&= {P^*}^2  + \big[ P^* (1-P^*)\nonumber\\
&  + (n-1) P^*\frac{1-p_1}{p_1-p_0} -  P^* \big ]  (p_1-p_0)^{n-1} \ .
\end{align}

\subsection{Calculation of $G^{ss}(n)$}

To compute $\Phi_n(S_0,S_n)$ and then $G^{ss}(n)$ we now consider the second step where $S_0$ and $S_n$ are relaxed. We define  $P(S_0,S_n | l_1, S_1 \ldots S_{n-1}, r_1 )$ as the probability of the state $\{S_0,S_n\}$ under the condition that the state  $\{ l_1, S_1 \ldots S_{n-1} ,r_1 \}$ has been reached after the first step (as indicated in Fig.~\ref{fig5}, $l_1$ and $r_1$ describe the states of the spins $S_{-1}$ and $S_{n+1}$, respectively). Knowing these probabilities and the probability of all  possible environments of $S_0$ and $S_n$ after the first relaxation step, we can write
\begin{widetext}
\begin{align}
 \Phi_n(S_0,S_n)& = 
      \sum_{l_1 r_1} {P^*}^{l_1+r_1} \left (1-P^* \right )^{2-l_1-r_1}\sum_{S_1\ldots S_{n-1}} P(S_1 \ldots S_{n-1}) \sum_{S_1\ldots S_{n-1}} P(S_1 \ldots S_{n-1})P(S_0,S_n | l_1, S_1 \ldots S_{n-1},r_1 ) .
\end{align}
\end{widetext}

If any of the spins $\{S_1,\ldots,S_{n-1} \}$ is up  the probability related to 
the second relaxation step is just a product of two independent terms. On the other hand, if all the spins $\{S_1,\ldots,S_{n-1} \}$ are down, the expressions of the probabilities are more  complicated since extra terms appear which account for the cases where the flip of $S_0$ (resp. $S_n$) triggers an avalanche which make all the spins $\{ S_1...S_{n-1} \}$ to flip up, changing the environment or the state of $S_n$ (resp. $S_0$). This yields
\begin{widetext}
\begin{align}
\small
\label{PS0Sn}
P(-1,-1 | l_1, S_1 \ldots S_{n-1}, r_1) & =  (1-p_{l_1+l_2}) (1-p_{r_1+r_2})  \nonumber   \\
P(+1,-1 | l_1, S_1 \ldots S_{n-1}, r_1) & =   
                     \left \{\begin{array} {lcr} 
                     p_{l_1+l_2} (1-p_{r_1+r_2})                                                                                       &   \mbox{if} & \{S_1\ldots S_{n-1}\} \neq \{-1\dots -1\} \\
                     p_{l_1} (1-p_{r_1}) - \left ( \frac{p_1-p_0}{1-p_0} \right )^{n-1}  p_{l_1} (p_{r_1+1}-p_{r_1}) \mbox{\hspace{2cm}} & \mbox{if} & \{S_1\ldots S_{n-1}\}  =  \{-1\dots -1\} \\ 
                            \end{array} \right.\nonumber  \\
P(-1,+1 | l_1, S_1 \ldots S_{n-1}, r_1) & =   
                     \left \{\begin{array} {lcr} 
                     (1-p_{l_1+l_2}) p_{r_1+r_2}                                                                                                     & \mbox{if} & \{S_1\ldots S_{n-1}\} \neq \{-1\dots -1\}   \\
                     (1-p_{l_1}) p_{r_1} -\left ( \frac{p_1-p_0}{1-p_0} \right )^{n-1} (p_{l_1+1}-p_{l_1}) p_{r_1}  \mbox{\hspace{2cm}}  & \mbox{if} & \{S_1\ldots S_{n-1}\}   =  \{-1\dots -1\} \\
                      \end{array} \right. \nonumber \\    
P(+1,+1 | l_1, S_1 \ldots S_{n-1}, r_1) & =   
                     \left \{\begin{array} {lcr} 
             p_{l_1+l_2} p_{r_1+r_2}                                                                                                               & \mbox{if} & \{S_1\ldots S_{n-1}\} \neq \{-1\dots -1\}  \\
             p_{l_1} p_{r_1} +\left ( \frac{p_1-p_0}{1-p_0} \right )^{n-1} \left[ p_{l_1}(p_{r_1+1}-p_{r_1})+ (p_{l_1+1}-p_{l_1}) p_{r_1} \right ] & \mbox{if} & \{S_1\ldots S_{n-1}\}  =   \{-1\dots -1\}  \\                             
                            \end{array} \right. \nonumber  \\                                        
\end{align}
\end{widetext}
\normalsize
where  $l_2=(1+S_{1})/2$ and $r_2=(1+S_{n-1})/2$.
Using Eqs.  (\ref{def_G}), (\ref{PS0Sn}), and the probability $ (1-p_0)^{n-1}$that all the spins $\{S_1\ldots S_{n-1} \}$ are down after the first relaxation step, we then obtain 
\begin{align}
\label{EqPn}
\Phi_n(-,-)& = \sum_{l_1,r_1} P(l_1,r_1)\nonumber\\
&\times \Big \{ \sum_{l_2,r_2} G_n(l_2,r_2) (1-p_{l_1+l_2}) (1-p_{r_1+r_2})  \Big \} \nonumber \\
\Phi_n(+,-)& =  \sum_{l_1,r_1} P(l_1,r_1)\Big \{  \sum_{l_2,r_2} G_n(l_2,r_2)  p_{l_1+l_2}  (1-p_{r_1+r_2})\nonumber\\
                    & - (p_1-p_0)^{n-1} p_{l_1} (p_{r_1+1}-p_{r_1} )  \Big \}    \nonumber \\                 
\Phi_n(-,+)& = \sum_{l_1,r_1} P(l_1,r_1) 
          \Big \{ \sum_{l_2,r_2} G_n(l_2,r_2) (1-p_{l_1+l_2})  p_{r_1+r_2}\nonumber\\
                  & - (p_1-p_0)^{n-1} (p_{l_1+1}-p_{l_1}) p_{r_1}  \Big \}       \nonumber \\ 
\Phi_n(+,+)& = \sum_{l_1,r_1} P(l_1,r_1)       
          \Big \{ \sum_{l_2,r_2} G_n(l_2,r_2) p_{l_1+l_2} p_{r_1+r_2}\nonumber\\
                    &+ (p_1-p_0)^{n-1} p_{l_1} (p_{r_1+1}-p_{r_1} )\nonumber\\
                   & + (p_1-p_0)^{n-1} (p_{l_1+1}-p_{l_1}) p_{r_1}  \Big \} .       
\end{align}
where  $P(l,r) \equiv {P^*} ^{l+r}  \left( 1-P^* \right )^{2-l-r}$. One can check that the following exact relations are satisfied:
\begin{align}
\label{Eqphi}
\Phi_n(+,-)=&\Phi_n(-+) \nonumber\\
\Phi_n(+,+)+2\Phi_n(+,-)+\Phi_n(-,-) =&1 \nonumber\\
\Phi_n(+,+)+\Phi_n(+,-)=&\frac{1}{2}(m(H)+1)
\end{align}
where the magnetization $m(H)$ is given by Eq. (\ref{Eqmag}).

\begin{figure}[hbt]
\begin{center}
\includegraphics[width=6cm,clip]{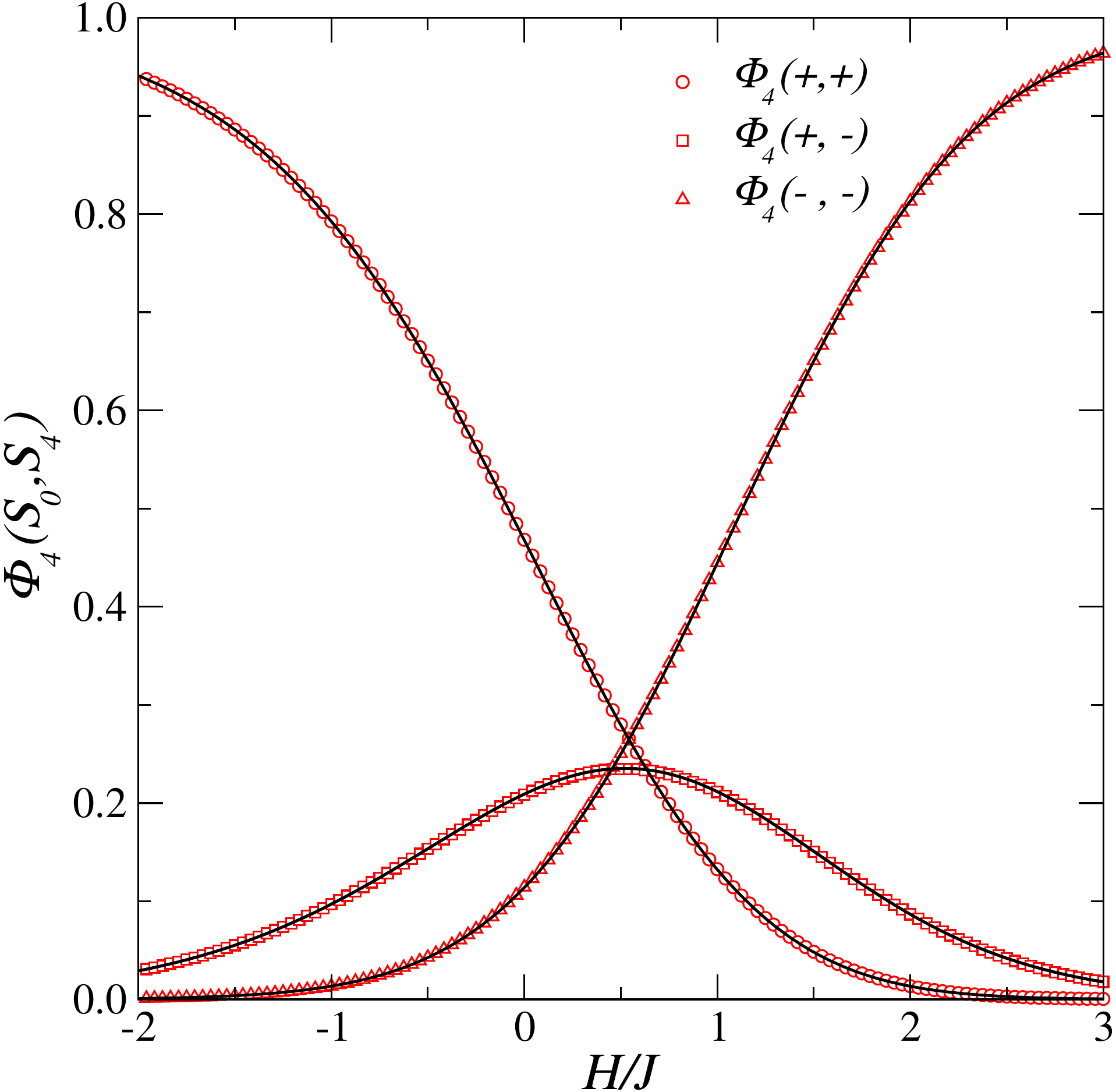}
 \caption{\label{fig6} Probabilities $\Phi_n(S_0,S_4)$ along the ascending branch of the hysteresis loop for $z=2$ and $\Delta=4$.  The simulation results (symbols) are compared to the predictions of Eqs. (\ref{EqPn}) (lines). Simulations were performed on random graphs with $N=10^6$ and the results were averaged over $1000$ disorder realizations.}
\end{center}
\end{figure}

In Fig.~\ref{fig6},  Eqs. (\ref{EqPn}) are compared to  simulation results in the case $n=4$. The excellent agreement confirms that the whole calculation is correct.

Note that  in Ref.\cite{KFS2000} it is stated  that the  probabilities $\Phi_n(S_0,S_n)$ are just linear combinations of the constrained probabilities $G_n$'s, {\it without} inhomogeneous terms.  Eqs. (\ref{EqPn}) show that this is only true for $\Phi_n(--)$. Finally, inserting Eqs. (\ref{EqPn}) in Eq. (\ref{EqS0Sn}) yields
\begin{align}
\overline{S_0 S_n}&=\sum_{l_1,r_1} P(l_1,r_1) \sum_{l_2,r_2} G_n(l_2,r_2)  (1-2p_{l_1+l_2})(1-2p_{r_1+r_2})\nonumber\\
&  +
4(p_1-p_0)^{n-1} \sum_{l_1,r_1} P(l_1,r_1) p_{l_1} (p_{r_1+1}-p_{r_1} )  \ .
\end{align}
A simpler expression is actually obtained by using Eqs. (\ref{Eqphi}) to express $G^{ss}(n)$ in terms of $\Phi_n(-,-)$ only, as done in Ref.\cite{KFS2000}. This  finally yields
\begin{align}
G^{ss}(n)&=4 \Big\{ \Phi_n(-,-) - [ \frac{1-m(H)}{2}]^2 \Big\} \nonumber\\
&= (p_1-p_0)^{n-1} [a + b(n-1)]
\end{align}
with
\begin{align}
a&=4P^*(Q^*-P^*) \left [  2(1-P^*)-P^*(Q^*-P^*) \right ]   \nonumber\\
b&=4P^*(Q^*-P^*)^2 \frac{(1-p_1)}{(p_1-p_0)} \ .
\end{align}
For $z=2$ we recall that
\begin{align}
P^*&=\frac{p_0}{1-p_1+p_0}   \nonumber\\
Q^*&=\frac{p_1-p_1^2+p_0p_2}{1-p_1+p_0} \ .
\end{align}

\subsection{Calculation of $G^{sh}(n)= \overline{S_0 h_n} $}

We now consider the spin-random field correlation function
\begin{equation}
\label{defGsh}
G^{sh}(n)\equiv\overline{S_0 h_n} = \sum_{S_0,S_n} \int dh_n S_0 h_n \Phi_n(S_0,S_n,h_n)
\end{equation}
where  $\Phi_n(S_0,S_n,h_n)$ is the probability density that the two spins $S_0$ and $S_n$ are in the state $\{S_0,S_n\}$ after  full relaxation of the system, with the random field acting on the spin $S_n$ having a value within $(h_n, h_n + dh_n)$. The calculation of these probabilities is  straightforward since we already
have  computed the probabilities $\Phi_n(S_0,S_n)$.  We only need to  restrict the integration of the random field distribution $\rho(h)$ over a range of $h$ compatible with  the state of the spin $S_n$. This gives

\begin{widetext}
\begin{align}
\Phi_n(-,-,h_n)& = \sum_{l_1,r_1} P(l_1,r_1)\Big \{\sum_{l_2,r_2} G_n(l_2,r_2)(1-p_{l_1+l_2}) \rho(h_n) \Theta [h_n<2(1-r_1-r_2)J-H]   \Big \}  \nonumber \\
\Phi_n(+,-,h_n)& = \sum_{l_1,r_1} P(l_1,r_1) \Big \{\sum_{l_2,r_2} G_n(l_2,r_2) p_{l_1+l_2} \rho(h_n) \Theta [h_n<2(1-r_1-r_2)J-H] \nonumber \\
         &  - (p_1-p_0)^{n-1} p_{l_1} \rho(h_n) \Theta [  2(1-r_1-1)J-H < h_n < 2(1-r_1)J -H ]    \Big \}                               \nonumber \\   
         \Phi_n(-,+,h_n)& = \sum_{l_1,r_1} P(l_1,r_1) \Big \{\sum_{l_2,r_2} G_n(l_2,r_2)(1-p_{l_1+l_2})\rho(h_n) \Theta [h_n>2(1-r_1-r_2)J-H] \nonumber \\
         & - (p_1-p_0)^{n-1} (p_{l_1+1} - p_{l_1}) \rho(h_n) \theta [h_n>2(1-r_1)J-H]     \Big \}                                          \nonumber \\       
\Phi_n(+,+,h_n)& = \sum_{l_1,r_1} P(l_1,r_1) \Big \{\sum_{l_2,r_2} G_n(l_2,r_2) p_{l_1+l_2} \rho(h_n)  \Theta [h_n>2(1-r_1-r_2)J-H]  \nonumber \\ 
     &+ (p_1-p_0)^{n-1} p_{l_1} \rho(h_n)  \Theta [  2(1-r_1-1)J-H < h_n < 2(1-r_1)J -H ]                                     \nonumber \\ 
     &  + (p_1-p_0)^{n-1} (p_{l_1+1} - p_{l_1}) \rho(h_n) \Theta [h_n>2(1-r_1)J-H]   \Big \}                            
\end{align}
\end{widetext}
where $\Theta(.)$ is the characteristic function of the domain indicated by the argument ({\it i.e.} $1$ inside the domain and $0$ outside). Inserting in Eq.(\ref{defGsh}) yields 
\begin{widetext}
\begin{align}
\overline{S_0 h_n} & =  \int dh_n h_n \big [ \Phi_n(+,-,h_n) + \Phi_n(+,+,h_n) - \Phi_n(-,-,h_n) -\Phi_n(-,+,h_n) \big ] \nonumber \\
                                   & =\sum_{l_1,r_1} P(l_1,r_1) \sum_{l_2,r_2} G_n(l_2,r_2)(2p_{l_1+l_2}-1)  \int_{-\infty}^{+\infty} dh_n h_n \rho(h_n)  \nonumber \\
                                   & +2(p_1-p_0)^{n-1} \sum_{l_1,r_1} P(l_1,r_1) (p_{l_1+1} -p_{l_1}) \int_{ 2(1-r_1)J-H}^{+\infty} dh_n h_n \rho(h_n) ,
\end{align}
\end{widetext}
and finally
\begin{align}
\label{Sohn}
\overline{S_0 h_n}&=2\Delta  (p_1-p_0)^{n-1}  
 \left [ \sum_{l_1} {P^*}^{l_1} (1-P^*)^{1-l_1} (p_{l_1+1} -p_{l_1}) \right ]\nonumber\\  
&\times\left[   \sum_{r_1} {P^*}^{r_1} (1-P^*)^{1-r_1} \rho\big(2(1-r_1)J-H\big) \right]  .
\end{align}

\section{Calculation of  $G^{ss}(2)$ and $G^{sh}(1)$ on the Bethe lattice.}

In this Appendix we  calculate $G^{ss}(2)$ and $G^{sh}(1)$ on a  Bethe lattice with coordination number $z$.
For completeness, we first recall the expressions of $G^{ss}(1)$ and $G^{sh}(0)$ obtained in Ref.\cite{IOV2005}:

\begin{equation}
\label{EqGss1}
G^{ss}(1)+m^2=1-4P^*+4P^*Q^*
\end{equation}
where $Q^*$ is given by
\begin{equation}
Q^{*}=\sum_{k=0}^{z-1} {z-1 \choose k} \left [ P^{*}(H) \right ]^k 
\left [1-P^{*}(H) \right ]^{z-1-k} p_{k+1}(H) \ ,
\end{equation}
and
\begin{align}
\label{EqGsh0}
G^{sh}(0)&=2\Delta\sum_{k=0}^z{z \choose k}\left [ p^{*}(H) \right ]^k\left [1-p^{*}(H) \right ]^{z-k} \nonumber\\
& \times\rho\big((z-2k)J-H\big) \ .
\end{align}

We recall that  $P^*$ (resp. $Q^ *$) is the probability that, along the ascending branch of the loop, a spin is up given that a neighbor is forced to be down (resp. up).

\subsection{Calculation of $G^{ss}(2)$}

\begin{figure}[hbt]
\begin{center}
\includegraphics[width=5cm,clip]{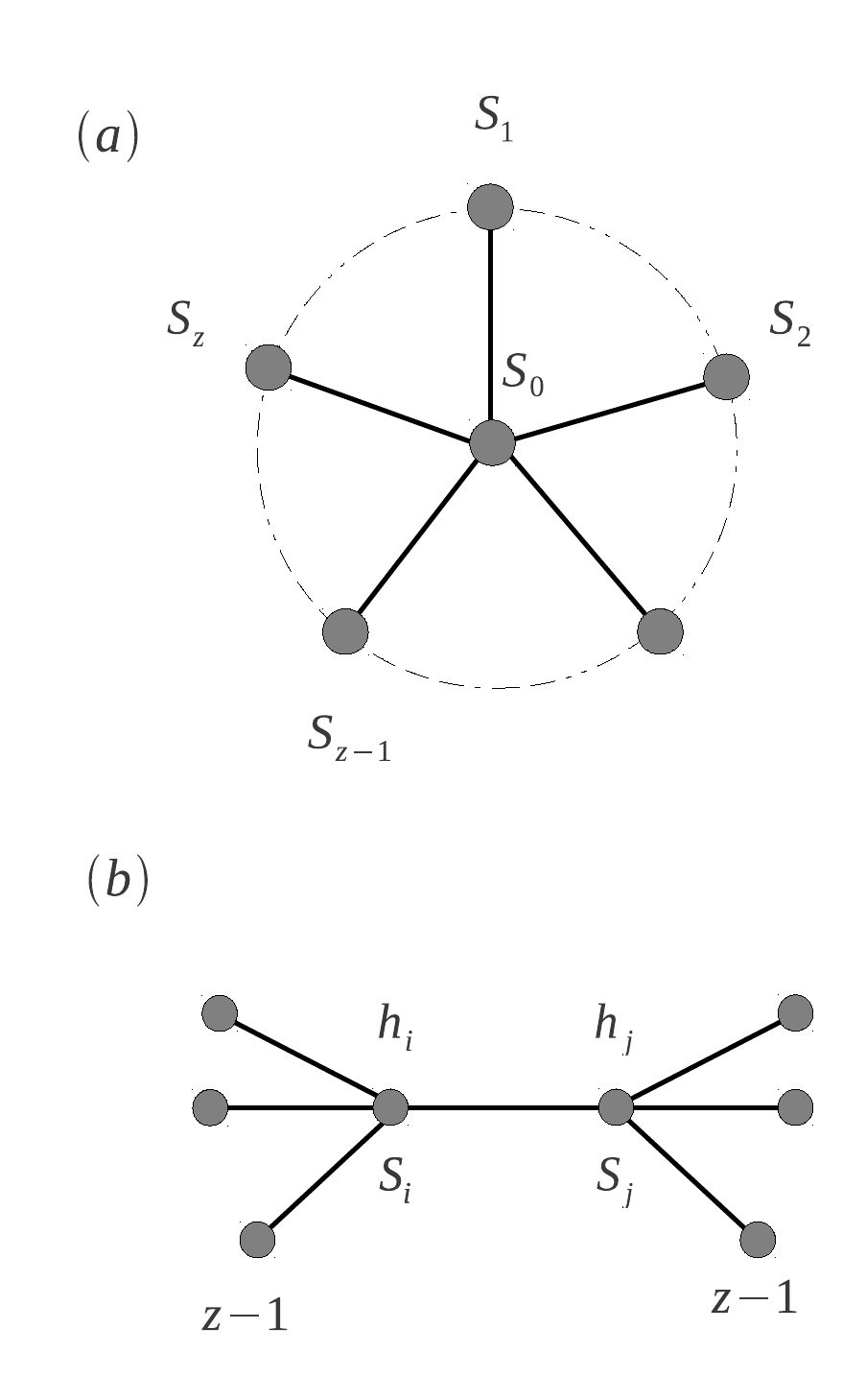}
 \caption{\label{fig7}Schematic representation of the environment of (a) a single spin $S_0$ and 
               (b) a pair of spins $S_i$ and $S_j$ .}
\end{center}
\end{figure}

To compute the correlations between  two spins $S_1$ and $S_2$ at the distance $n=2$ we  consider a central 
spin $S_0$ and its $z$ neighbors $ \{ S_1,\ldots S_z  \} $ as depicted in Fig. \ref{fig7}a. 
By definition,
\begin{align}
\overline{ S_1 S_2 } &= \sum_{S_0,S_1 \ldots S_z} S_1 S_2 P(S_0,S_1 \ldots S_z) \nonumber\\
& =  \sum_{S_1,S_2} S_1 S_2 \big[\sum_{S_3\ldots S_z} P(-1,S_1 \ldots S_z) \nonumber\\
&+\sum_{S_3 \ldots S_z} P(+1,S_1 \ldots S_z) \big ] \nonumber\\
&=\sum_{S_1,S_2} S_1 S_2 \left [ P(-1,S_1,S_2)+ P(+1,S_1,S_2) \right ]
\end{align}
where $P(S_0,S_1,\ldots,S_z)$ is the probability of having the configuration $\{S_0,S_1,\ldots,S_z\}$ 
when the system is fully relaxed.

By relaxing the spins in two steps, we obtain 
\begin{align}
\label{eq1}
P(-1,S_1\dots  S_z) &= \left [ 1-P^{*} \right ]^{z-q}  \left [ P^{*} \right ]^q  (1-p_q)  \\
\label{eq2}
P(+1,S_1\ldots S_z) &= \left [ 1-Q^{*} \right ]^{z-q} \sum_{k=0}^{q} {q \choose k}\nonumber\\
&\times\left [ Q^{*}-P^{*} \right ]^{q-k} \left [ P^{*} \right ]^k  p_k 
\end{align}
where $q$ is the number of neighbors of $S_0$ that are up when the system is fully relaxed,
\begin{equation}
q=\frac{\sum_{j=1}^{z} S_j + z}{2} = 0,1, \ldots z. 
\end{equation}
Eq.~(\ref{eq1}) is rather straightforward (recall that $P^ *$ is the probability that a spin is up given that a neighbor is forced to be down). In Eq.~(\ref{eq2}), the summation accounts for all the different ways of having $S_0=1$ when $q$ neighbors are up, and the term in front of the summation accounts for the probability that  $S_0$ has $z-q$ of his neighbors down  if it is up. 

Using Eqs. (\ref{eq1}) and (\ref{eq2}) we find
\begin{align}
P(-1,S_1,S_2) &=\sum_{l=0}^{z-2} {z-2 \choose l}  \left [ 1-P^{*} \right ]^{z-l-n}  \left [ P^{*} \right ]^{l+n}  (1-p_{l+n}) \nonumber \\
P(+1,S_1,S_2) &=\sum_{l=0}^{z-2} {z-2 \choose l}  \left [ 1-Q^{*} \right ]^{z-l-n} \Big [\sum_{k=0}^{l+n} {l+n \choose k} \nonumber\\
&\times \big [ Q^{*}-P^{*} \big]^{l+n-k} \left [ P^{*} \right ]^k p_k \Big]
\end{align}
where
\begin{equation}
n=\frac{S_1+S_2 + 2}{2} = 0,1,2
\end{equation}
We finally obtain the following  expression for the correlations 
at the next-nearest neighbor distance
\begin{align}
\label{EqGss2}
\overline{ S_1 S_2 } &=
\sum_{n=0}^{2} (-1)^n {2 \choose n} \sum_{l=0}^{z-2} {z-2 \choose l} \nonumber\\
&\times \Bigg \{  \left [ 1-P^{*} \right ]^{z-l-n}  \left [ P^{*} \right ]^{l+n} (1-p_{l+n}) \nonumber\\
&+ \left [ 1-Q^{*} \right ]^{z-l-n} \left[ \sum_{k=0}^{l+n} {l+n \choose k} \left [ Q^{*}-P^{*}\right ]^{l+n-k} \left [ P^{*} \right ]^k p_k \right]  \Bigg \}
\end{align}

\subsection{Calculation of $G^{sh}(1)$}

To compute
\begin{equation}
G^{sh}(1)=\overline{ S_i h_j } =
\sum_{S_i} \int dh_j S_i h_j P(S_i,h_j) \ ,
\end{equation}
we again relax the spins in two steps so to obtain 
\begin{align}
\label{EqPsihj}
P(S_i,h_j) &=
\sum_{l=0}^{z-1} {z-1 \choose l} {P^*}^l (1-P^*)^{z-1-l} \nonumber\\
&\times\sum_{r=0}^{z-1} {z-1 \choose r} {P^*}^r (1-P^*)^{z-1-r}
P(S_i,h_j | l , r),
\end{align}
where $P(S_i,h_j | l , r)$ is the probability that the spin at $i$ is in the state $S_i$ after the second relaxation step, with the random field acting on $S_j$ having a value within $(h_j, h_j + dh_j)$, and under the condition that
the environment of $S_i$ and $S_j$ is in the state $(l,r)$ after the first relaxation step (see Fig.~\ref{fig7}b).
 As in Ref.\cite{IOV2005}, $l=1,\ldots, z-1$ (resp. $r=1,\ldots, z-1$) is the number of neighbors of $S_i$ ($S_j$) that are up, without taking into account $S_j$ (resp. $S_i$).

These probabilities are given by
\begin{align}
P(-1,h_j| l,r)& =
\left\{ 
\begin{array}{lrr}
    (1-p_l) \rho(h_j) & \mbox{if} & h_j<(z-2r)J-H \\
(1-p_{l+1}) \rho(h_j) & \mbox{if} & h_j>(z-2r)J-H 
\end{array}
\right. \nonumber \\
P(+1,h_j| l,r)& =
\left\{ 
\begin{array}{lrr}
    p_l     \rho(h_j) \mbox{\hspace{0.9cm}}& \mbox{if} & h_j<(z-2r)J-H \\
    p_{l+1} \rho(h_j) \mbox{\hspace{0.9cm}}& \mbox{if} & h_j>(z-2r)J-H  \ .
\end{array}
\right.
\end{align}

Inserting these expressions in  Eq. (\ref{EqPsihj}) and after some algebra we finally obtain
\begin{align}
\label{EqGsh1}
G^{sh}(1)&=2\Delta
\left[\sum_{l=0}^{z-1} {z-1 \choose l} {P^*}^l (1-P^*)^{z-1-l} (p_{l+1}-p_l) \right] \nonumber\\
&\times\left[\sum_{r=0}^{z-1} {z-1 \choose r} {P^*}^r (1-P^*)^{z-1-r} 
\rho ((z-2r)J-H) \right] \ .
\end{align}
For $z=2$ and  $n=1$, one can check  that this equation gives back Eq.~(\ref{Sohn}).


\begin{thebibliography}{10}
\bibitem{SDP2006} J. P. Sethna, K. A. Dahmen, and O. Perkov\'ic in {\it The Science of Hysteresis}, edited by G. Bertotti and I. Mayergoyz, Acedemic Press, Amsterdam (2006).
\bibitem{SDKKRS1993} J. P. Sethna, K. A. Dahmen, S. Kartha, J. A. Krumhansl, B. W. Roberts, and J. D. Shore, Phys. Rev. Lett. 70, 3347 (1993).
\bibitem{DKRT2005} F. Detcheverry, E. Kierlik, M. L. Rosinberg, and G. Tarjus, Phys. Rev. E 72, 051506 (2005). 
\bibitem{B2008} F. Bonnet, T. Lambert, B. Cross, L. Guyon, F. Despetis, L. Puech, and P. E. Wolf, Europhys. Lett. 82, 56003 (2008).
\bibitem{DSS1997} D. Dhar, P. Shukla, and J.P. Sethna, J. Phys. A {\bf 30}, 5259 (1997); S. Sabhapandit, P. Shukla, and D. Dhar, J. Stat. Phys. {\bf 98}, 103 (2000); P. Shukla, Phys. Rev. E {\bf 63}, 027102 (2001).
 \bibitem{CGZ2002} F. Colaiori, A. Gabrielli, and S. Zapperi, Phys. Rev. B {\bf 65}, 224404 (2002).
 \bibitem{ABCDDMZ2005} M. J. Alava, V. Basso, F. Colaiori, L. Dante, G. Durin, A. Magni, and S. Zapperi, Phys. Rev. B {\bf 71}, 064423 (2005).
\bibitem{IOV2005} X. Illa, J. Ort\'in, and E. Vives, Phys. Rev. B {\bf 71}, 184435 (2005); X. Illa, P. Shukla, and E. Vives, Phys. Rev. B, {\bf 73} 092414  (2006).
\bibitem{OS2010} H. Ohta and S. Sasa, Euro. Phys. Lett. {\bf 90}, 27008 (2010).
\bibitem{RT2010} M.L. Rosinberg and G. Tarjus, J. Stat. Mech. P12011  (2010).
\bibitem{H2004} For a recent review, see E. Hoinkis, Part. Part. Syst. Charact. 21, 80 (2004).
\bibitem{DKRT2006} F. Detcheverry, E. Kierlik, M. L. Rosinberg, and G. Tarjus, Phys. Rev. E {\bf 73}, 041511 (2006).
\bibitem{KFS2000} J. C. Kimball, H. L. Frisch, and L. Senapati, Physica A {\bf 279}, 151 (2000).
\bibitem{N1998} See {\it e.g.} T. Natterman, {\it Spin glasses and random fields} (World Scientific, Singapore, 1998).
\bibitem{S1996} P. Shukla, Physica A {\bf 233}, 235 (1996).
\bibitem{GM1983} G. Grinstein and D. Mukamel, Phys. Rev. B {\bf 27}, 4503 (1983).
\bibitem{LN1989} J.M. Luck and Th.M. Nieuwenhuizen, J. Phys. A {\bf 22}, 2151 (1989); see also J.M. Luck, {\it Syst\`emes D\'esordonn\'es Unidimensionnels} (Saclay Al\'ea, 1992). On the other hand,  in the random one-dimensional lattice-gas  studied by Y. Fonk and H. J. Hilhorst, J. Stat. Phys. {\bf 49}, 1235 (1987),  the correlation function at $T=0$ also displays the $n$ prefactor.
\bibitem{FLM2001} D. Fisher, P. Le Doussal, and C. Monthus, Phys. Rev. E {\bf 64}, 066107 (2001). In this work, the nonequilibrium dynamics of the RFIM chain after a quench from a random initial condition ({\it e.g.} a high temperature state) is studied via an asymptotically exact real space renormalization group analysis. As a byproduct, the equilibrium quantities at low temperature are obtained in the long time limit, at the end of the renormalization procedure.
\bibitem{IM1975} Y. Imry and S. K. Ma, Phys. Rev. Lett. {\bf 35}, 1399 (1975).
\bibitem{KW1981} H. S. Kogon and D. J. Wallace, J. Phys. A: Math. Gen. {\bf 14} L527 (1981).
\bibitem{HPT2011} T.P. Handford, F-J Perez-Reche, and S. N. Taraskin, arXiv:cond-mat/1106.3424v1.
\bibitem{HM2006} J. P. Hansen and I. R. McDonald, {\it Theory of simple liquids} (Academic Press, 2006).
\bibitem{note1} However this type of `Ornstein-Zernike' approximation implies that the small-wavevector behavior of the correlation function is regular so that the anomalous dimension $\eta$ is zero.
\bibitem{HS1977} J. S. Hoye and G. Stell, J. Chem. Phys. {\bf 67}, 439 (1977); Mol. Phys. {\bf 52}, 1071 (1984); D. Pini and G. Stell, Phys. Rev. Lett. {\bf 77}, 996 (1996); D. Pini, G. Stell, and R. Dickman, Phys. Rev. E {\bf 57}, 2862 (1998).
\bibitem{GKRT2001} S. Grollau, E. Kierlik, M.L. Rosinberg and G. Tarjus, Phys. Rev. E {\bf 63}, 041111 (2001); S. Grollau, M.L. Rosinberg and G. Tarjus, Physica A {\bf 296}, 460 (2001).
\bibitem{KRT1997} E. Kierlik, M. L. Rosinberg, and G. Tarjus, J. Stat. Phys. {\bf 89}, 215 (1997); {\it ibid} 
{\bf 94}, 805 (1999); {\it ibid} {\bf100}, 423 (2000).
\bibitem{G1994} A. Giacometti, J. Phys. A: Math. Gen. {\bf 28}, L13 (1995).
\bibitem{MT1996} C. Monthus and C. Texier, J. Phys. A: Math. Gen. {\bf 29}, 2399 (1996).
\bibitem{note2} This is especially true in the three-dimensional RFIM since the anomalous dimensions $\eta$ and ${\overline \eta}$ associated to the connected and disconnected correlation functions, respectively, are rather large. 


\end{thebibliography}
\end{document}